\documentclass[aps,prl,reprint,groupedaddress,preprintnumbers]{revtex4-2}

\usepackage{tikz}
\RequirePackage{amssymb,amsmath}
\RequirePackage{hyperref}

\usepackage{url}

\newcommand{\be}{\begin{equation}}
\newcommand{\ee}{\end{equation}}

\newcommand{\nn}{\nonumber}
\newcommand{\bea}{\begin{eqnarray}} 
\newcommand{\eea}{\end{eqnarray}}
\newcommand{\bse}{\begin{subequations}}	
\newcommand{\ese}{\end{subequations}}
\newcommand{\bal}{\begin{align}} 
\newcommand{\eal}{\end{align}}

\newcommand{\bi}{\begin{itemize}} 
\newcommand{\ei}{\end{itemize}}

\def\a{\alpha}    \def\d{\delta} 
 
  \def\h{\eta} \def\k{\kappa}
\def\l{\lambda}  \def\m{\mu}
\def\n{\nu} \def\o{\omega}   \def\r{\rho}
\def\s{\sigma}

\begin{document}

\preprint{RBI-ThPhys-2021-38}

\title{Axion gravitodynamics, Lense-Thirring effect, and gravitational waves}

\author{Athanasios Chatzistavrakidis}
\email{Athanasios.Chatzistavrakidis@irb.hr}
\affiliation{Division of Theoretical Physics, Rudjer Bo\v skovi\'c Institute, Bijeni\v cka 54, 10000 Zagreb, Croatia}
\affiliation{Erwin Schrödinger International Institute for Mathematics and Physics,
Boltzmanngasse 9, A-1090, Vienna, Austria}
\author{Georgios Karagiannis}
\email{Georgios.Karagiannis@irb.hr}
\author{George Manolakos}
\email{Georgios.Manolakos@irb.hr}

\affiliation{Division of Theoretical Physics, Rudjer Bo\v skovi\'c Institute, Bijeni\v cka 54, 10000 Zagreb, Croatia}

\author{Peter Schupp}
\email{p.schupp@jacobs-university.de}
\affiliation{Department of Physics and Earth Sciences, Jacobs University Bremen,  28759 Bremen, Germany}

\begin{abstract}
We investigate physical implications of a gravitational analog of axion electrodynamics with a parity-violating gravitoelectromagnetic theta term. This is related to the Nieh-Yan topological invariant in gravity with torsion, in contrast to the well-studied gravitational Chern-Simons term quadratic in curvature, coupled via a dynamical axionlike scalar field. Axion gravitodynamics is the corresponding linearized theory. We find that potentially observable effects are over 80 orders of magnitude stronger than for its Chern-Simons counterpart and could be in reach for detection by experiments in the near future. For a near-Earth scenario, we derive corrections to the Lense-Thirring effect and compare them to data from satellite-based experiments (Gravity Probe B). For gravitational waves, we find modified dispersion relations, derive the corresponding polarization-dependent modified group and phase velocities, and compare them to data from neutron star mergers (GW170817) to derive even stronger bounds.
\end{abstract}

\maketitle

\section{Introduction}

General Relativity and its modifications or alternative theories of gravity are being tested in our times beyond the classical experimental tests that were settled in the 20th century. Prominent physical contexts that generate such tests are gravitational waves and corrections to the precession of gyroscopes orbiting rotating bodies such as the Earth. Both are predictions of General Relativity and both have been corroborated in the 21st century through the detection and measurements of the LIGO and Gravity Probe~B experiments respectively \cite{LIGOScientific:2016aoc,Everitt:2011hp}.

The linearized limit of General Relativity plays a central role in both cases above. Notably, Einstein's equations in this weak field limit are similar to Maxwell's equations for electrodynamics, with the electric and magnetic field replaced by the gravitational and the gravitomagnetic field. The latter is generated by rotating masses and it is the backbone of the Lense-Thirring effect~\footnote{Various phenomenological aspects of the effect have been discussed in \cite{Iorio:2010rk}.}. Having this analogy in mind, a natural question regards the existence and the potential physical effects of a gravitational theta term, the gravitational counterpart of the electromagnetic term $\theta \vec{E}\cdot \vec{B}$ in the Lagrangian. The latter, in case $\theta$ is not the same everywhere, appears in axion electrodynamics \cite{Wilczek:1987mv} and it has important physical uses, notably as an effective field theory in the physics of topological materials \cite{Qi:2010qag,Sekine:2020ixs}, such as topological insulators and superconductors \cite{Qi:2008ew,Qi:2012cs}. 

Axion Gravitodynamics is the theory of linearized gravity that contains the gravitational theta term $\theta \vec{E}_{G}\cdot \vec{B}_{G}$, with the appearance of the gravitational field $\vec{E}_{G}=-\vec\nabla\Phi-\frac{1}{2}\dot{\vec{A}}$ and the gravitomagnetic field $\vec{B}_{G}=\vec{\nabla}\times\vec{A}$, in terms of the Newton and gravitomagnetic potentials $\Phi$ and $\vec{A}$ (we drop the subscript G in the notation of the fields from now on until the conclusion, since we will always refer to the gravitational and never to the electromagnetic ones). Remarkably, this term has been used in the recent past to study the gravitational response of topological materials \cite{Nomura:2011hn,Furusaki:2012zf,Parrikar:2014usa,Liang2019TopologicalME}. 
Its nonlinear origin was proposed in~\cite{Chatzistavrakidis:2020wum} to be the teleparallel equivalent of General Relativity with the Weitzenb\"ock connection, modified with the Nieh-Yan term \cite{Nieh:1981ww}, the quadratic in torsion topological invariant in four dimensions \cite{Li:1999ue,Chandia:1997hu}.
It is worth noting that the Nieh-Yan term has found numerous applications in recent years, for instance in modified Einstein-Cartan \cite{Shaposhnikov:2020frq} or teleparallel gravity \cite{Mielke:2009zz} and in cosmological scenarios \cite{Mielke:2006zp,Langvik:2020nrs} and moreover it is directly related to the  Holst term that has played a pivotal role in loop quantum gravity \cite{Ashtekar:1986yd,Mercuri:2007ki,Mercuri:2009zi,Banerjee:2010yn,Taveras:2008yf}.  

Our purpose in this paper is to study the effect of axion gravitodynamics on the precession of gyroscopes and on the propagation of gravitational waves. In a similar spirit to the Peccei-Quinn mechanism in QCD, we promote the parameter $\theta$  to an axion-like field $\theta(t,\vec{x})$ \footnote{Torsional axions were also studied in Refs. \cite{Castillo-Felisola:2015ema,Hohmann:2020dgy}.}. This introduces a new dimensionful scale, essentially the decay constant of the axion-like field, which should be bounded by observations. 

The setting is analogous to the Chern-Simons modification of General Relativity \cite{Jackiw:2003pm}, which was used in~\cite{Smith:2007jm} to calculate corrections to the general-relativistic gravitomagnetic field around massive spinning bodies in case $\theta$ is only time-dependent (as in the quintessence scenario). Although a parity-violating interaction between the gravitoelectric and gravitomagnetic fields is introduced in both settings \footnote{An interesting discussion on parity violation and torsion in (quantum) gravity appears in \cite{Freidel:2005sn}}, the crucial difference is that the Chern-Simons modification is based on higher derivatives (since the curvature and torsion tensors differ by one derivative) and therefore an $\vec{E}\cdot \vec{B}$ term is not sufficient for a nontrivial coupling to $\theta$. On the contrary, in axion gravitodynamics it is precisely the $\vec{E}\cdot \vec{B}$ term that couples to $\theta$, which leads to different quantitative predictions. What is more, we argue that the effect of this term in axion gravitodynamics is much stronger and therefore if it exists it becomes much more relevant than its Chern-Simons counterpart. Comparing with the Gravity Probe~B experiment, we find that the coupling is very weak and therefore it should be either detected in upcoming experiments
or certain reasonable models will be ruled out.      

In a similar spirit to the modification induced by the electromagnetic theta term to the propagation of light~\cite{Carroll:1989vb}, we find that the dispersion relation of gravitational waves is modified in axion gravitodynamics. This leads to a splitting of the group and phase velocities of the two polarizations of the gravitational wave, which is a direct consequence of  parity and Lorentz violation. Using the observed discrepancy between the group velocity of the gravitational wave event GW170817 and the speed of light~\cite{LIGOScientific:2017zic}, we are able to place an additional bound on the coupling of the gravitational theta term. This bound is stricter than the one obtained previously by five orders of magnitude.

\section{Axion Gravitodynamics}

The most common way to obtain the Einstein field equations of General Relativity from an action principle is to consider the Einstein-Hilbert action, which is a second-order formulation with the only independent field being the Riemannian metric. However, there exists a variety of different starting points that also lead to the Einstein equations (see e.g. the textbook \cite{Ortin:2004ms} for a detailed exposition.). For example one may consider as independent variable the vierbein (or tetrad) or employ the first-order (Palatini) formulation where the independent variables are the metric and the linear connection, respectively the vierbein and the spin connection. 

Einstein's General Relativity may also be viewed as a particular case in a class of gravity theories formulated in terms of the Weitzenb\"ock connection, which has vanishing curvature but nonvanishing torsion \cite{Ortin:2004ms}. The independent field in this case is the vierbein $e^{a}{}_{\mu}$, where $\mu$ and $a$ are spacetime and tangent (Lorentz) indices respectively, both ranging from 0 to 3; the torsion of the Weitzenb\"ock connection is not an independent field but it is given by $T_{\mu\nu}^{~~\rho}=2\,e_{a}{}^\rho\partial_{[\mu}e_{~\nu]}^{a}$ in terms of the vierbein and its inverse. The correspondence to the usual formulation of General Relativity is based on the identity \footnote{This immediately follows from an identity relating the Riemann tensor for the Levi-Civita connection and the contorsion tensor \cite{Ortin:2004ms,Hayashi:1979qx}.}
\be \label{identity}
\frac{1}{16}R^{\text{LC}}=\mathbb{T}-\frac{1}{4}\nabla_\mu T^{\m\n}{}_\n\,,
\ee
which relates the Ricci scalar $R^{\text{LC}}$ of the Levi-Civita connection to the torsion of the Weitzenb\"ock connection up to a total derivative term. The scalar $\mathbb{T}$ is the following linear combination of the three parity-preserving, quadratic in the first derivative of the vierbein terms called Weitzenb\"ock invariants:
\be \label{scalar combo}
 \mathbb{T}=\frac{1}{4}T_{\mu\nu\rho}T^{\mu\nu\rho}+\frac{1}{2}T_{\mu\nu\rho}T^{\rho\nu\mu}-T_{\nu\mu}{}^{\mu} T^{\nu\rho}{}_{\rho}\,. 
\ee 
Due to the identity \eqref{identity}, the Einstein-Hilbert action (written in terms of the vierbein) is identical, up to boundary terms, to the action functional 
\be \label{Stegr}
S_{\text{TEGR}}=-\frac{1}{2\k^2}\int \mathrm{d}^4x\,e\,\mathbb{T}\,,
\ee 
where $\k^2=8\pi G$ and $e$ is the vierbein determinant. The two formulations are therefore classically equivalent. In fact, this action is the starting point for the so-called teleparallel equivalent formulation of General Relativity.

Jackiw and Pi considered a modification of General Relativity, extending it by a parity-violating gravitational Chern-Simons term which is quadratic in the Riemann curvature tensor \cite{Jackiw:2003pm}. Subsequently, the authors of \cite{Smith:2007jm} studied the effects of this modified theory on the precession of gyroscopes orbiting around the Earth, placing bounds on the allowed parameter space of the theory. A crucial ingredient for this analysis was the coupling of the Chern-Simons term to a pseudoscalar axion-like field. Motivated by these works we pose the following question. Suppose that instead of the usual formulation of General Relativity in terms of the Levi-Civita connection which has curvature but not torsion, we start with the equivalent formulation in terms of the Weitzenb\"ock connection which has torsion but not curvature. How can the theory be modified by a gravitational Chern-Simons-like term that is quadratic in the torsion tensor and what physical consequences can be obtained once an axion-like field is coupled to the theory? 

The answer to this question is simple once one notices that there exists a four-dimensional, parity-violating and Lorentz invariant topological invariant which is quadratic in the torsion tensor. This is the Nieh-Yan $4$-form defined in \cite{Nieh:1981ww} for an arbitrary connection. In a local coordinate basis, the components of this $4$-form read  
\be \label{NY}
 \frac{1}{12}\,{\cal N}_{\m\n\r\s}:= \frac{1}{2}\,T_{[\m\n}{}^\l T_{\r\s]\l}-R_{[\m\n\r\s]}
\ee
in terms of the components of the curvature and torsion tensors. Furthermore, it was shown in Ref. \cite{Nieh:1981ww} that this object corresponds to a total derivative due to the following identity:
\be 
 \frac{1}{12}\,{\cal N}_{\m\n\r\s}=\partial_{[\m}T_{\n\r\s]}\,.
\ee
In absence of torsion (e.g. for the Levi-Civita connection) the Nieh-Yan $4$-form \eqref{NY} vanishes identically due to the algebraic Bianchi identity for the Riemann tensor. 
On the other hand, choosing the curvature-free but torsion-full Weitzenb\"ock connection, one is left with a single torsion-squared term. 

Based on these observations, a modification of General Relativity formulated in terms of the Weitzenb\"ock connection is singled out; it is obtained by adding the quadratic torsion term of Eq. \eqref{NY} to the action functional \eqref{Stegr}. Since this is a total derivative term, only a coupling to an additional (pseudo)scalar field $\theta(x)$ would lead to modified dynamics, as is the case in the Chern-Simons modified gravity too. Such axion-like fields appear in a variety of Standard Model extensions as pseudo-Nambu-Goldstone bosons of spontaneously broken global symmetries and they are also abundant in models arising from string theory compactifications as dark matter, inflaton or quintessence candidates \cite{Ringwald:2014vqa,Choi:2020rgn}. It is then natural to assume that if such a field exists it can couple to the quadratic torsion term. Therefore we consider the action functional (we work in units where $c=1$)
\bea \label{nonlinearaction}
   S&=& \frac{1}{2\kappa^2}\int \mathrm{d}^4x\,e\left(-\mathbb T+\frac{\ell\theta(x)}{4}\epsilon^{\mu\nu\rho\sigma}T_{\mu\nu\lambda}
 T_{\rho\sigma}^{~~\lambda}\right)\nn\\ &&  
   - \int \mathrm{d}^4x\,e \left(\frac{1}{2}(\partial\theta)^2+V(\theta)\right)+S_{\text{M}}\,, \label{action}
\eea
where $V(\theta)$ is the potential of the field $\theta(x)$. Since $\theta(x)$ has mass dimension 1, a scale $\ell$ with dimensions of length is introduced in the theory.

It is important to note that although both the Chern-Simons modified gravity and the Weitzenb\"ock-based teleparallel gravity modified by the Nieh-Yan term are equivalent to General Relativity in absence of the field $\theta$ and its coupling to topological terms, the modified theories are not equivalent to each other. Crucially, this additional term is quadratic in derivatives in the latter case, in contrast to Chern-Simons modified gravity where it is quartic in derivatives. Since in the theory given by \eqref{nonlinearaction} there are less derivatives in the term that modifies the dynamics compared to Chern-Simons gravity, this is expected to lead to stronger physical effects and therefore the theory we introduced deserves a separate study.

The field equations obtained from this theory after variation with respect to the vierbein and the field $\theta$ are the modified Einstein  and  Klein-Gordon equations 
\bea \label{EinsteinEq}
 G_{\mu\nu}+C_{\mu\nu}= \kappa^2 T_{\mu\nu}\,,\, \square\,\theta =\frac{dV}{d\theta}-\frac{\ell}{8\k^2}\epsilon^{\mu\nu\rho\sigma}T_{\mu\nu\lambda} T_{\rho\sigma}^{~~\lambda}\,.\nn
\eea 
Here $G_{\m\n}$ is the usual Einstein tensor and $C_{\m\n}$ is obtained from the variation of the Nieh-Yan ($\theta$-)term with respect to the vierbein, 
\be 
    C_{\mu\nu}={\ell}\epsilon_{\mu\k\rho\sigma} T^{\rho\sigma}_{~~\nu}\partial^\k\theta\,.
\ee
As usual, $T_{\m\n}$ is the total energy-momentum tensor containing contributions both from the matter fields and from $\theta(x)$. 
Inspection of the modified Einstein equation leads to two remarks: First, the antisymmetric part of $C_{\m\n}$ vanishes, which can be seen as the on-shell constraint $\epsilon^{\lambda\rho\sigma[\mu}T_{\rho\sigma}^{~~\nu]}=0$ for the torsion tensor. Second, one may confirm that the divergence of $C_{\m\n}$ is equal to the divergence of the energy-momentum tensor for $\theta(x)$, which is a useful consistency check. 

In the present paper we are interested in the weak field limit of this theory, which we are now going to derive. Since we are not going to delve in the dynamics of $\theta$, hence we consider it effectively nondynamical and treat it as a background field.
In the weak field limit, the vierbein and its inverse are split into flat background values and perturbations as $e^a{}_\mu\simeq \d^a_\mu+a^a{}_\mu$ and $e_a{}^\mu\simeq \d_a^\mu-a_a{}^\mu$, where $a^a{}_\mu=\delta^{a\n}a_{\n\m}$ is a small perturbation containing both a symmetric $h_{\mu\nu}:=2a_{(\mu\nu)}$ and an antisymmetric $b_{\mu\nu}:=2a_{[\mu\nu]}$ piece. The former corresponds to the linearized metric perturbation around the flat Minkowski metric, since $g_{\mu\nu}=e^a{}_\mu e^b{}_\nu \eta_{ab}= \eta_{\mu\nu}+2a_{(\mu\nu)}+\mathcal{O}(a^2)$, while the latter is the linearized Kalb-Ramond field. In this limit, the torsion tensor becomes $T_{\m\n\r}\simeq \partial_{[\m}h_{\n]\r}+\partial_{[\m}b_{\n]\r}$. Then the first line of the nonlinear action \eqref{nonlinearaction} reduces to the linearized one 
\bea \label{linearizedaction}
   \hspace{-0.2cm}S&\simeq& -\frac{1}{8\kappa^2}\int \mathrm{d}^4x\,\left(2\,\partial_\m h_{\n\r}\partial^{[\m}h^{\n]\r}+2\,\partial_\m h\,\partial_\n h^{\m\n}\right.\nn\\ && 
   -\partial_\m h\,\partial^\m h-\partial_\mu h^{\mu\nu}\partial^\rho h_{\rho\nu}-\ell\,\theta \,\epsilon^{\m\n\r\s}\partial_\m h_{\n\l}\partial_\r h_{\s}^\l
   \nn\\ && 
   \left.-2\,\ell\,\theta \,\epsilon^{\m\n\r\s}\partial_\m b_{\n\l}\partial_\r h_{\s}^\l -\ell\,\theta \,\epsilon^{\m\n\r\s}\partial_\m b_{\n\l}\partial_\r b_{\s}{}^\l\right),
\eea
where we denote the trace of the linearized metric by $h:= \eta_{\m\n}h^{\m\n}$. This action is invariant under linearized diffeomorphisms parametrized by an arbitrary vector $\xi^\m$, which act on the fields as 
\be \d h_{\m\n}=\partial_{\m}\xi_{\n}+\partial_{\nu}\xi_{\mu} \quad  \text{and} \quad  \d b_{\m\n}=\partial_{\m}\xi_{\n}-\partial_{\nu}\xi_{\mu}\,.
\ee
In the following we will be interested in backgrounds where the Kalb-Ramond perturbation vanishes, $b_{\m\n}=0$. Then, the field equations obtained by variation of the action \eqref{linearizedaction} with respect to $h_{\m\n}$ are 
\bea \label{lin EinstEqs}
&&\Box \,h_{\m\n}-\eta_{\m\n}\Box\, h+\partial_\mu\partial_\n h-2\partial^\r \partial_{(\m}h_{\n)\r}\nn\\ && 
+\eta_{\m\n}\partial_\r\partial_\s h^{\r\s}-\ell\,\epsilon_{\k\r\s(\m}\partial^\k\theta \,\partial^\r h^\s_{\n)}=-\k^2T_{\m\n}\,.
\eea
One can then perform a coordinate transformation and reach the harmonic gauge $\partial^{\m}\bar{h}_{\m\n}=0$, where $\bar{h}_{\m\n}:=h_{\m\n}-\frac{1}{2}\h_{\m\n}h$ is the trace-reversed linearized metric. Then the field equations take the simpler form 
\be \label{AG equation}
\Box\,\bar h_{\m\n}-\ell\,\varepsilon_{\rho\sigma\lambda(\m}\partial^\rho\theta\,\partial^\sigma \bar h_{\n)}^\lambda=-\k^2\,T_{\m\n}\,.
\ee
Taking the divergence of the linearized field equation in vacuum, we observe that it is consistent with the harmonic gauge condition provided the second derivatives of $\theta(x)$ are negligible. The energy-momentum tensor for a finite distribution of matter has the form  $T_{\mu\nu}=\rho\,v_\mu v_\nu$, where $\rho$ is the mass density of the source. We will use the notation $J_i= \rho v_i$ for the components of the mass current vector with Latin indices denoting spatial directions.

The theory given by the field equations \eqref{AG equation} is what we call axion gravitodynamics. To see why, recall that within the  framework of gravitoelectromagnetism (see e.g. the review \cite{Mashhoon:2003ax}) scalar and vector potentials are defined through the components of $\bar{h}_{\m\n}$ as
\be 
\Phi:=-\frac{1}{4}\bar{h}_{00}\quad  \text{and} \quad A_{i}:=\frac {1}{2}\bar{h}_{i0}\,.
\ee 
Subsequently, one can define the gravitational and gravitomagnetic fields by $\vec{{E}}:=-\vec{\nabla}\Phi-\frac{1}{2}\dot{\vec{A}}$ and $\vec{ B}:=\vec{\nabla}\times\vec{ A}$. In terms of these, the linearized field equations \eqref{AG equation} take the form
\begin{eqnarray}
     \label{GEM equations}
     &&    \vec{\nabla}\cdot\vec{E}=-\frac{\k^2}{2}(\rho+\rho_{\theta})\,,\quad \vec{\nabla}\times\vec{E}+\frac{1}{2}\dot{\vec{B}}=\vec{0}\,,\\
  &&  \vec{\nabla}\cdot\vec{B}=0\,, \quad \vec{\nabla}\times\Vec{B}-2\dot{\vec{E}}=-\k^2(\vec{J}+\vec{J}_{\theta})\,,\nn
\end{eqnarray}
where the contributions of $\theta(x)$ can be seen as the effective mass and mass current densities 
\be \rho_{\theta}=\frac{\ell}{\k^2}\vec{\nabla}\theta\cdot\vec{B}\quad\text{and}\quad \vec{J}_{\theta}=-\frac{\ell}{\k^2}\left(\vec{\nabla}\theta\times\vec{E}+\frac{\dot\theta }{2}\vec{B}\right)
\ee 
on the right hand side of the field equations.
In the coming sections we will solve these equations in specific physical settings. It is worth mentioning that formally they have the same form as the ones found by Sikivie in the context of axion models \cite{Sikivie:1983ip} and by Wilczek in axion electrodynamics \cite{Wilczek:1987mv}.

\section{Effect on Gyroscope Precession}

We would now like to use axion gravitodynamics to compute the gravitomagnetic field of a spinning spherical body, for instance a planet, in case $\theta$ is spatially homogeneous and slowly-varying, such as a quintessence field \cite{Carroll:1998zi}-\footnote{The general proposal of using an axion as quintessence appears in \cite{Kim:2002tq}, although in a different scenario to ours.}. 
We further assume that the source is stationary, i.e.  $\dot{\phi}=\dot{A}^i=\dot{E}^i=0$. Under these assumptions, one can act on the fourth equation in \eqref{GEM equations} with the operator $\left(\frac{\ell \,\dot\theta}{2} \mathbb I+\vec{\nabla}\times \right)$ and, using $\vec{\nabla}\cdot\vec{B}=0$, obtain 
\be \label{final equation}
\left(\nabla^2+\frac{\ell^2\dot\theta^2}{4}\right)\vec{B}=\k^2\left(\frac{\ell\,\dot\theta}{2} \vec{J}+ \vec{\nabla}\times\vec{J}\right)\,.
\ee
This is an inhomogeneous Helmholtz equation for the gravitomagnetic field $\vec{B}$. To solve it, we consider the case of a homogeneous spherical source of radius $R$ with constant density $\rho$ and angular velocity $\vec{\o}$. The mass current of such a matter distribution is given by
\be \label{J}
\vec{J}(\vec{r})=-\rho \,r\, \Theta(R-r)\, \hat{r}\times\vec{\o}\,.
\ee
The general solution of equation \eqref{final equation} for such a source reads as
\be \begin{split}\label{final B}
\vec{B}&=\k^2\rho R^2\left[f_1(r)\,\vec\o+f_2(r)\,\hat{r}\times\vec\o+f_3(r)\,\hat{r}\times(\hat{r}\times\vec\o)\right]\,,\nn
\end{split}
\ee
in terms of the functions
\be \label{f out}
\begin{split}
f_1(r)&=\frac{2R}{r}\,j_2(\frac{\ell\,\dot\theta}{2} R)\,y_1(\frac{\ell\,\dot\theta}{2} r)\,,\\
f_2(r)&=-\frac{\ell\,\dot\theta}{2} R\,j_2(\frac{\ell\,\dot\theta}{2} R)\,y_1(\frac{\ell\,\dot\theta}{2} r)\,,\\
f_3(r)&=\frac{\ell\,\dot\theta}{2} R\,j_2(\frac{\ell\,\dot\theta}{2} R)\,y_2(\frac{\ell\,\dot\theta}{2} r)
\end{split}
\ee
for the exterior region ($r\geq R$). The elementary functions $j$ and $y$ correspond to spherical Bessel functions of the first and second kind, respectively.
This solution is continuous on the boundary of the sphere and in the GR limit $\ell\to 0$ it reproduces the well-known result for the gravitomagnetic field generated outside a rotating spherical body 
\be 
\vec{B}_{GR}= -\k^2\rho R^2\left[\frac{2R^3}{15r^3}\,\vec\o+\frac{R^3}{5r^3}\,\hat{r}\times(\hat r\times\vec\o)\right]\,.
\ee

This result highlights an important qualitative feature of the gravitational $\theta$ term, which was also found in Chern-Simons modified gravity in Ref.~\cite{Smith:2007jm}. The gravitomagnetic field has an additional component parallel to the vector $\hat{r}\times\vec\omega$. This toroidal component is absent in GR and appears as a consequence of the parity violation introduced by the topological terms in each case.   

Following the analogous discussion in Ref.~\cite{Smith:2007jm}, we can now compute the correction induced by the $\theta$ term in the Schiff precession of a gyroscope \cite{doi:10.1119/1.1935800}, while it performs a circular polar orbit around a rotating spherical body. Using the standard relation 
\be \label{general relation}
\dot{\vec{S}}=2\vec{B}\times\vec{S}\,,
\ee 
between the time variation of the spin $\vec{S}$ of the gyroscope and the gravitomagnetic field generated by the spherical body, one can show  that the ratio $\a:=|\langle\dot{\vec{S}}_{\theta}\rangle|/|\langle\dot{\vec{S}}_{GR}\rangle|$ reads as
\be \label{fraction}
    \a=1+\frac{15r^2}{R^2}\,j_2(\frac{\ell\,\dot\theta}{2} R)\left(y_1(\frac{\ell\,\dot\theta}{2} r)+\frac{\ell\,\dot\theta}{2} r\,y_0(\frac{\ell\,\dot\theta}{2} r)\right),
\ee
where $r$ and $R$ are the radii of the gyroscopic orbit and the spherical body, respectively.  $\dot{\vec{S}}_\theta$ is the contribution of $\theta$ to the change in gyroscope's spin and  $\dot{\vec{S}}_{GR}$ the standard change predicted by GR, averaged over one orbital period. 
This result is qualitatively similar to the one obtained in Ref.~\cite{Smith:2007jm} for Chern-Simons modified gravity, $\a_{\text{CS}}$.
However, the two ratios $\a$ and $\a_{\text{CS}}$ are quantitatively different.

\begin{figure}[htbp]
\centerline{\includegraphics[scale=.5]{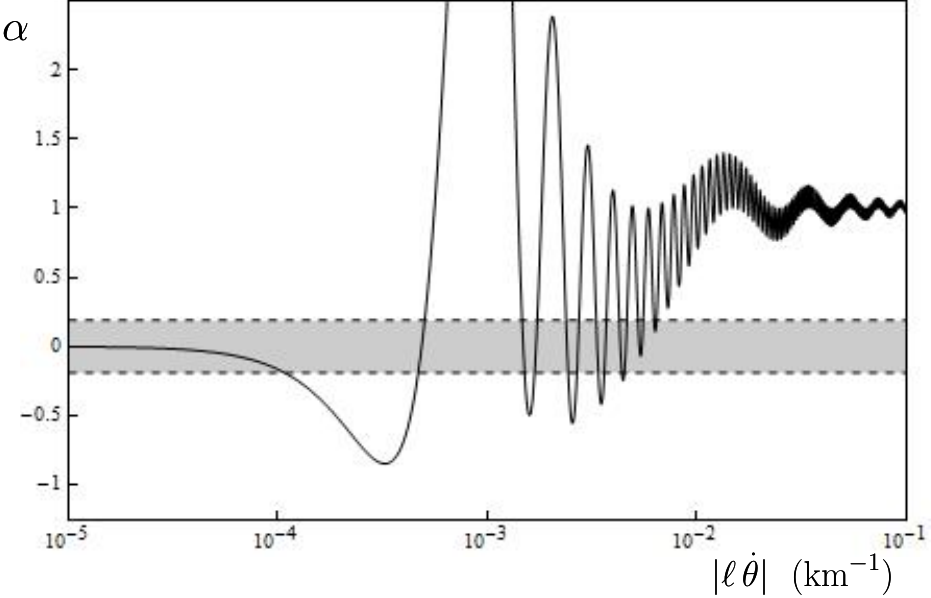}}
\caption{\emph{The ratio $\a$ in \eqref{fraction} for the Gravity Probe B mission. The $x$-axis is logarithmic in the variable $|\ell\,\dot\theta|$. The gray region enclosed by the dashed lines $\a=\pm0.19$ represents the $19\%$ accuracy with which the Gravity Probe B final results confirm General Relativity.}}
\label{fig}
\end{figure}

Measurement of the Schiff precession for gyroscopes orbiting the Earth at an altitude of $642\,\text{km}$ was one of the primary goals of the Gravity Probe B mission \cite{Everitt:2011hp}. The final results of the mission reported a verification of the general-relativistic result at an accuracy of $19\%$. The laser-ranged satellites LAGEOS, LAGEOS~2, and LARES further improve the accuracy to about $5\%$ \cite{LageosLares}, but here we will work with the more conservative result. %%%%
If we assume that the Earth is a perfect sphere with radius $R=6368\,\text{km}$, then the satellite's orbit has a radius $r=7010\,\text{km}$. Using these values, we can then plot the ratio \eqref{fraction} as depicted in Figure \ref{fig}. 
From this plot we conclude that a $19\%$ verification of the general-relativistic effect places the following generic upper bound:
\be \label{bound GPB}
|\ell\,\dot\theta| \; \lessapprox \; 10^{-4}\,\text{km}^{-1}\simeq2\times  10^{-23}\,\text{GeV}\,.
\ee 
One should also note that for certain non generic values of the fundamental parameters it can be relaxed by 1-2 orders of magnitude. 
It is useful to compare the strengths of the predicted effects between Axion Gravitodynamics and Chern-Simons modified gravity. In the latter case and in terms of the same parameters $\ell$ and $\dot{\theta}$, Ref.\cite{Smith:2007jm} find
\be 
|\ell_{CS}\,\dot\theta|\lessapprox 10^{60}\,\text{GeV}\,.
\ee
Thus we observe that the bound obtained from axion gravitodynamics is immensely stricter that the Chern-Simons one. 
For a field 
$\theta(t)=\theta_0 e^{-t/\ell}$ and, therefore $\ell\,\dot\theta\approx\theta$, the above bounds imply that the coupling of the topological term in axion gravitodynamics is weaker than the Chern-Simons one by approximately $83$ orders of magnitude. The bound \eqref{bound GPB} will be further improved in the next section.

\section{Effect on Gravitational Waves}

Let us now study the gravitational wave solution of the system of equations \eqref{GEM equations}, for the case of a slowly varying and spatially homogeneous axion field. For gravitational waves propagating in vacuum, we also have to consider the source-free case of vanishing $\rho$ and $\vec{J}$. Acting on the fourth equation in \eqref{GEM equations} with a time derivative, neglecting second time derivatives of $\theta$, and using the second equation in \eqref{GEM equations} to eliminate $\dot{\vec{B}}$, leads to
\be 
\Box\vec{E} 
=-\frac{\ell\,\dot\theta}{2}\,\vec{\nabla}\times\vec{E}\,.
\ee
This PDE is solved by a transversal circular polarized wave $\vec{E}$ with frequency $\o$, wave vector $\vec{k}$ and dispersion relation 
\be\label{dispersion relation}
\omega^2-k^2=\pm \frac{\ell\dot\theta}{2}\,k\,,\ee
where $k\equiv |\vec{k}|$. The positive and negative signs correspond to left and right helicity respectively. For $k>\frac{\ell\dot\theta}{2}$, these dispersion relations have real solutions. The requirement $k>\frac{\ell\dot\theta}{2}$ holds by default for gravitational waves, since their typical wavelengths are $\lambda_{GW}=\frac{2\pi}{k}\simeq10^3\text{km}\simeq 5\times10^{21}\text{GeV}^{-1}$ and $|\ell\,\dot\theta|$ should respect the bound \eqref{bound GPB}.

Let us now discuss the consequences of the modified dispersion relation \eqref{dispersion relation} to the phase and group velocities of a gravitational wave. These are found to be
\be \label{velocities}\begin{split}
v_p&=\frac{\omega}{k} = \sqrt{1\pm \frac{\ell\dot\theta}{2k}}\simeq 1\pm\frac{\ell\dot\theta}{4k}+\mathcal{O}(\ell^2)\,,\\
v_g&=\frac{d\omega}{dk}=\frac{1\pm\frac{\ell\dot\theta}{4k}}{\sqrt{1\pm\frac{\ell\dot\theta}{2k}}}
\simeq 1+ \frac{\ell^2\dot\theta^2}{32 k^2}+\mathcal{O}(\ell^3)\,,
\end{split}\ee
respectively. The front velocities ($k \rightarrow \infty$ limit) are 1, i.e.\ equal to the speed of light.
Similarly to the electromagnetic case of \cite{Carroll:1989vb}, we observe that the two polarizations propagate with different velocities, which are also different from the speed of light. The group velocity is larger than the speed of light and polarization-independent up to $\mathcal{O}(\ell^2)$. These are direct consequences of the parity and Lorentz violation, respectively, which are introduced through the gravitational theta term. Moreover, we observe that the group velocity is modified at order $\mathcal{O}(\ell^2)$, while the phase velocity is already affected at order $\mathcal{O}(\ell)$.

The group velocity of the neutron star merger gravitational wave event GW170817 was bounded by observation to be \cite{LIGOScientific:2017zic} 
\be \label{observation bound}
-3\times 10^{-15}\,\leq\, v_g -1\,\leq\,7\times 10^{-16}.
\ee
This is a bound on the ratio of the speed of gravitational waves over the speed of light.  
The group velocity \eqref{velocities} predicted by axion gravitodynamics can be compared to the observed one \eqref{observation bound}. As we already mentioned, the typical wavelengths of gravitational waves are $\lambda_{GW}\simeq 5\times10^{21}\text{GeV}^{-1}$. Ignoring terms of order $\mathcal{O}(\ell^3)$, we can therefore obtain the upper bound
\be \label{final bound}
|\ell\,\dot\theta| \; \lessapprox \; 2 \times 10^{-28}\,\text{GeV}\,.
\ee
This is stricter than the one in \eqref{bound GPB} by  five orders of magnitude. We discuss this further in the conclusions.

\section{Conclusions}

Axion gravitodynamics is the theory that incorporates in the weak field limit of General Relativity a gravitational $\theta$~term proportional to the product  $\vec{E}_G\cdot\vec{B}_G$ of the gravitational and gravitomagnetic fields. It arises naturally as the linearization of the teleparallel equivalent of General Relativity with a Nieh-Yan term. Promoting $\theta$ to a field and introducing an associated length scale $\ell$, leads to a set of gravitational field equations for $\vec{E}_G$ and ${\vec{B}_G}$ that include first derivatives of $\theta(x)$ in striking similarity to axion electrodynamics. Motivated by the variety of uses of the latter in physics, it is natural to ask what is the effect of axion gravitodynamics in certain  physical phenomena, such as the Lense-Thirring effect and the propagation of gravitational waves in vacuum.

In this paper, under the assumption that the field $\theta$ is spatially homogeneous and depends only on cosmic time, we solved the gravitational field equations of the theory in these two settings. First we computed the gravitomagnetic field around a spinning spherical mass and its corrections to the general relativistic prediction. At a qualitative level, we found additional components to the gravitomagnetic field with no analogue in General Relativity, which were first reported in the study of Chern-Simons modified gravity \cite{Smith:2007jm}. Comparing with the observational data of the Gravity Probe B satellite mission, a bound on the parameters of the theory is found, $|\ell\,\dot\theta| \, \lessapprox\, 2\times  10^{-23}\,\text{GeV}$, which is much stronger than the one in the Chern-Simons case (\,$\lessapprox\, 2\times  10^{60}\,\text{GeV}$). Therefore, we conclude that the effect of axion gravitodynamics is stronger and if it exists it could very well be discovered or ruled out by near future observations \cite{Kehl:2016mgp,Sakstein:2017pqi,Tartaglia:2021idn}. 

The presence of $\theta(t)$ has an effect on the propagation of gravitational waves too, leading to a modification of their dispersion relation which influences the phase and group velocities of gravitational waves in a polarization-dependent way (birefringence).
Comparison to the reported bounds on the group velocity from the neutron star merger GW170817 allowed us to place an even stricter bound to the parameters of axion gravitodynamics, $|\ell\,\dot\theta| \,\lessapprox\, 2 \times 10^{-28}\,\text{GeV}$. Together with the Lense-Thirring effect, the modified gravitational wave propagation can serve as observational signatures of Axion Gravitodynamics. 

Although we have not discussed the dynamics of the field $\theta$, appealing to its (quintessential) axionic nature prompts us to interpret this as a bound on its decay constant $|f_\theta|>10^{28}|\dot\theta|\,\text{GeV}$. 
The decay constant for a quintessence field should be approximately equal to (and definitely not higher than) the reduced Planck mass $M_P\simeq 3\times 10^{18}\,\text{GeV}$ where quantum gravitational effects become relevant. Assuming that $f_\theta\lesssim M_P$ directly implies that $|\ell|\gtrsim M_P^{-1}$. This would mean that $|\dot\theta|\lesssim 6\times10^{-10}\text{GeV}^2$, a reasonable outcome given that to explain dark energy the kinetic energy of quintessence should be much lower than its potential energy. More detailed study on the dynamics of $\theta$ is required to test these statements more precisely. Finally, it would be interesting to explore the possibility of a field that is not spatially homogeneous. In that case, finding solutions to the field equations becomes more demanding and we plan to report on this in future work. 

\

\paragraph{Note added.} Recently, we learned of the preprint \cite{Wu:2021ndf}, which has some overlap with our discussion on gravitational waves and a proposal for a stronger bound on the parameters of the theory based on phase velocities.  

\

\begin{acknowledgments} 
We would like to thank Chris Hull, Alekos Kehagias and Claus Lämmerzahl for useful discussions. This work is supported by the Croatian Science Foundation Project ``New Geometries for Gravity and Spacetime'' (IP-2018-01-7615). We express our gratitude to the DFG Research training Group 1620 ``Models of Gravity''. A.\ Ch. is grateful to the Erwin Schr\"odinger International Institute for Mathematics and Physics for hospitality and financial support in the framework of the Research in Teams Project: Higher Global Symmetries and Geometry in (non-)relativistic QFTs. P.\ S. thanks the Nordic Institute for Theoretical Physics (Nordita) for hospitality. 
\end{acknowledgments}

\bibliography{theta-term_gravity}

\end{document}